# Valorizing the carbon byproduct of methane pyrolysis in batteries

Ji, Y.,[a,b] Palmer, C.,[c] Foley, E. E.,[a,b] Giovine, R.,[a,b] Yoshida, E.,[a,b]
McFarland, E.,[c] and Clément, R. J.[a,b*]

a. Materials Department, University of California Santa Barbara, California 93106, United States
b. Materials Research Laboratory, University of California Santa Barbara, California 93106, United States
c. Chemical Engineering Department, University of California Santa Barbara, California 93106, United States

* Corresponding author. Tel: 805-893-4294. E-mail: rclement@ucsb.edu (Raphaële J. Clément)

## Abstract

While low-cost natural gas remains abundant, the energy content of this fuel can be utilized without greenhouse gas emissions through the production of molecular hydrogen and solid carbon via methane pyrolysis. In the absence of a carbon tax, methane pyrolysis is not economically competitive with current hydrogen production methods unless the carbon byproducts can be valorized. In this work, we assess the viability of the carbon byproduct produced from methane pyrolysis in molten salts as high-value-added anode or conductive additive for secondary Li-ion and Na-ion batteries. Raman characterization and electrochemical differential capacity analysis demonstrate that the use of molten salt mixtures with catalytically-active $FeCl_3$- or $MnCl_2$ result in more graphitic carbon co-products. These graphitic carbons exhibit the best electrochemical performance (up to 272 mAh/g of reversible capacity) when used as Li-ion anodes. For all carbon samples studied here, disordered carbon domains and retained salt species trapped and/or intercalated into the carbon structure were identified by X-ray photoelectron and multinuclear solid-state nuclear magnetic resonance spectroscopy. The latter lead to reduced electrochemical activity and reversibility, and poorer rate performance compared to commercial carbon anodes.

The electronic conductivity of the pyrolyzed carbons is found to be highly dependent on their purity, with the purest carbon exhibiting an electronic conductivity nearly on par with that of commercial carbon additives. These findings suggest that more effective removal of the salt catalyst could enable applications of these carbons in secondary batteries, providing a financial incentive for the large-scale implementation of methane pyrolysis for "low-carbon" hydrogen production.

1. Introduction

Over the next several decades, society will need to lessen its reliance on fossil resources and adopt significantly more sustainable practices. Hydrogen is a vital chemical intermediate and may play an important role in our long-term energy infrastructure. Hydrogen is a clean, energy-dense fuel[1] produced globally at a rate of 80-100 million metric tons (MT) a year,[2] predominantly via steam methane reforming (SMR; $CH_4 + H_2O \rightarrow CO + 3H_2$) followed by a water-gas shift (WGS; $CO + H_2O \rightarrow CO_2 + H_2$) reaction. Such processes result in a significant amount of carbon dioxide gas byproduct and are energy intensive.[3]

An alternative method to produce hydrogen is through the thermal decomposition of methane.[3] Unlike SMR (plus WGS), methane pyrolysis (MP; $CH_4 \rightarrow C + 2H_2$) results in a solid carbon product that is environmentally benign and more readily stored and utilized than $CO_2$. Fundamentally, MP requires less energy input per unit of hydrogen produced than SMR+WGS; however, it requires twice as much methane. Today, in the absence of a meaningful carbon tax hydrogen production via MP is more expensive than current methods that produce $CO_2$, and several comparative technoeconomic analyses have concluded that an alternative way to financially incentivize hydrogen production via MP is to valorize the carbon byproduct.[4–6]

High temperature methane pyrolysis is technically challenging and the use of molten metals or molten salts as both a heat transfer and reaction medium has been reported by multiple research groups.[4,7–14] The advent of such molten metal/salt media allows the carbon product to be continuously separated from the reactor, unlike conventional solid catalyst configurations in which the carbon is inseparable and causes coking within the reactor.[7,13,15] While the utilization and valorization of these molten-media-derived carbon products is often discussed in the literature, very few works have evaluated their performance in technologically-relevant applications. One of the few examples is the work by Tang et al.[16], who produced graphitic nanosheets from MP in a molten copper environment and demonstrated their effectiveness in sorbent and electrical heating applications. Some studies have also explored the decomposition of waste plastics in molten salt environments and assessed the properties (such as conductivity) of the residual carbonaceous materials.[17,18]

Some of the more profitable and large-scale applications of carbons are in secondary batteries. For instance, lithium-ion batteries (LIBs) have become the technology of choice for powering portable electronic devices and electric vehicles, and for renewable energy storage. Current devices contain several carbon-based components, the main one being the graphitic anode, which accounts for 10 to 15% of the total raw material cost.[19] Technoeconomic analyses indicate that a carbon value on the order of 200 $/MT would render MP economically competitive with the SMR+WGS process.[20] Battery grade synthetic graphite costs about 20,000 $/MT, while the natural graphite analogue costs at least 8,000 $/MT, with market reports suggesting that demand for graphite for Li-ion anode applications will increase six-fold by 2030.[21] Although less profitable than Li-ion anodes, carbonaceous conductive additives represent about 5 to 20% of the total weight of cathode composites[22] and are valued between 1000 and 2000 $/MT.[23] Hence, even considering the lower end of these values, the ability to produce carbons of sufficient quality to be used as anode materials or conductive additives would render MP competitive with the SMR+WGS process. Furthermore, the implementation of carbons produced from MP in commercial Li-ion cells would lower raw material cost and largely meet market demands for battery grade graphitic carbon.

With this in mind, we examine the structure and evaluate the electronic conductivity and electrochemical properties of carbons produced via MP in Li-ion and Na-ion cells. The carbons are obtained using a differential molten salt bubble column under an Ar environment at a temperature greater than 1000°C and using one of five molten salt catalysts: NaBr, NaCl, KCl, Na-KCl (with 3 wt.% $FeCl_3$), and KCl: $MnCl_2$ (1:1 ratio). These carbons are hereafter referred to as NaBr carbon, NaCl carbon, KCl carbon, $FeCl_3$ carbon, and $MnCl_2$ carbon, respectively. Using a combination of X-ray photoelectron spectroscopy (XPS) and solid-state nuclear magnetic resonance (ssNMR) spectroscopy, we show that residual catalytic salt is present in all of the carbon byproducts (despite a hot water wash) and resides as either physically trapped salt-like units or as co-intercalated salt cations and anions. Raman and differential capacity analysis in Li-ion cells highlight the impact of the molten salt used in the MP process on the carbon structure. Specifically, molten salt mixtures containing $MnCl_2$ or $FeCl_3$ produce carbon that is more graphitic in character, while MP in simple alkali halide salts produces carbons that are more disordered, agreeing with previous findings.[7,13,14] Interestingly, in carbons produced in molten salt mixtures containing $MnCl_2$ or $FeCl_3$, residual salt impurities are electrochemically active and contribute to the capacity observed in Li-ion cells. The $MnCl_2$ carbon electrode exhibits the best electrochemical performance, with a high capacity of 272 mAh/g in a Li-ion half-cell, yet it suffers from a relatively low rate performance that is attributed in part to the presence of salt impurities. The electronic conductivity of the pyrolyzed carbons also strongly depends on their purity, with the higher purity (NaCl) carbon exhibiting a conductivity of about 80-85% of that of standard carbon conductive additives. Our findings suggest that the use of MP carbon byproducts in battery devices is conceivable and, with further improvements in carbon purification processes, would financially incentivize hydrogen production.

## 2. Materials and Methods

### 2.1. Carbon synthesis and washing steps

The carbons were obtained using a differential molten salt bubble column under an Ar environment at a temperature greater than 1000°C using one of five molten salt catalysts: NaBr, NaCl, KCl, Na-KCl (with 3 wt.% $FeCl_3$), and KCl:$MnCl_2$ (1:1 ratio). The carbon products were separated from the melt and cooled to room temperature. The hot water washing step consisted of soaking the carbon in hot water at 70°C for 2 hours with constant stirring. The additional $NO_2BF_4$ washing step consisted of soaking the hot water washed carbon in 0.1 M $NO_2BF_4$ in acetonitrile solution for 1 day with constant stirring. The washed carbon powders were dried at 130°C for 72 hours under vacuum and stored in an Ar-packed glovebox before characterization.

### 2.2. XPS characterization

High resolution XPS spectra were obtained on a Kratos Axis Ultra XPS system on hot water washed KCl and NaBr carbon powders. All reference salts used were purchased from Sigma Aldrich and dried before the measurement. The NaBr reference salt was kept in an inert atmosphere at all times during storage and transfer.

### 2.3. NMR characterization

$^{23}$Na, $^{79}$Br, and $^{13}$C solid-state NMR data were collected on hot water washed NaBr carbon powder at a magnetic field $B_0 = 18.8$ T (800 MHz for $^1$H) using a standard bore Bruker Biospin spectrometer equipped with an AVANCE-III console. The spectra were obtained using a 2.5 mm double-resonance HX magic angle spinning (MAS) probe tuned to $^{23}$Na (211.65 MHz), $^{79}$Br (200.46 MHz), or $^{13}$C (201.19 MHz). Around 8.5 mg of the dry carbon powder was packed in a 2.5 mm zirconia rotor and closed with a Vespel® cap in an argon-filled glovebox. A spacer made of PTFE tape was placed between the carbon powder and the Vespel® cap to (i) improve air tightness and (ii) equilibrate the rotor. To further avoid moisture exposure, the rotor was spun at $\nu_R = 30$ kHz using dry nitrogen. $^{23}$Na and $^{79}$Br chemical shifts were referenced against 1 M aqueous NaCl and KBr salt solution at 0 ppm. $^{13}$C chemical shifts were referenced against tetramethylsilane using the CH$_2$ resonance of pure adamantane as a secondary reference ($\delta_{iso}(^{13}C) = 38.5$ ppm). Solid-state NMR data were processed using the Bruker TopSpin 3.6.0 software and spectra were fitted using DMfit software.[24]

The $^{23}$Na spin echo spectrum was obtained using a rotor synchronized spin echo sequence (90°−$\tau_R$−180°−$\tau_R$) and 90° and 180° radio-frequency (RF) pulses of 1.7 and 3.4 µs, respectively. Data were averaged over 512 scans and collected with a recycle delay of 5 s to ensure full relaxation of the $^{23}$Na signals. An isotropic $^{23}$Na NMR spectrum was recorded using the projected magic angle turning phase adjusted sideband separation (pj-MATPASS) pulse sequence, which effectively removes spinning sidebands due to MAS.[25,26] This experiment used a 90° RF pulse of 1.7 µs, a recycle delay of 2 s, and data were averaged over 100 scans. A $^{23}$Na triple-quantum (3Q-MAS) 2D experiment was also carried out using a z-filtered pulse sequence.[27–30] Excitation and reconversion pulses lasted 5 and 2 µs, respectively, while the last RF pulse consisted of a central transition selective 90° pulse of 10.5 μs. Data were averaged over 3600 scans with a recycle delay of 0.5 s leading to an experimental time of 18 h. The sheared 2D spectrum was obtained using the "xfshear" procedure in TopSpin 3.6.0.

The $^{79}$Br spin echo spectrum was obtained using a rotor synchronized spin echo sequence (90°–$\tau_R$–180°–$\tau_R$) and 90° and 180° RF pulses of 0.9 and 1.8 μs, respectively. Data were averaged over 4096 scans and collected with a recycle delay of 0.1 s to ensure full relaxation of the $^{79}$Br signals.

The $^{13}$C spin echo spectrum was obtained using a rotor synchronized spin echo sequence (90°–$\tau_R$–180°–$\tau_R$) and 90° and 180° RF pulses of 1.9 and 3.8 μs, respectively. Data were averaged over 2880 scans with a recycle delay of 30 s leading to an experimental time of 24 h. To ensure steady-state condition prior to the acquisition of the $^{13}$C NMR data, a train of RF pulses was applied before the recycle delay to saturate the $^{13}$C magnetization. This train used 50 RF pulses of 1.8 μs, with a 5 ms delay between consecutive pulses.

### 2.4. Electrochemical testing of carbon electrodes

Electrodes were prepared by hand mixing the carbon powder with a carbon Super P conductive additive (MTI Corp.) in an 85:5 weight ratio for 30 minutes. The composite was then hand-ground with polytetrafluoroethylene (PTFE, Sigma Aldrich) for 10 mins to make a thick electrode with an 85:5:10 weight ratio of pyrolyzed carbon. The thick electrode was hand-rolled into a film to prepare a thin electrode with a loading density of 5 mg/cm$^2$. All electrochemical tests against Li were conducted in Swagelok-type cells with a 200 μL 1 M LiPF$_6$ in 1:1 w/w EC:DMC (Sigma Aldrich) electrolyte with less than 25 ppm in water content, a lithium metal (Alfa Aesar) counter electrode, and a glass fiber separator (Whatman GF/D). All electrochemical tests against Na were conducted using a 1M NaPF$_6$ in 1:1 w/w EC:DMC electrolyte (prepared in-house) with less than 25 ppm in water content. All electrochemical tests were carried out galvanostatically under a current rate of C/20 based on 1 Li$^+$ intercalation per 6 C atoms (corresponding to the theoretical capacity of graphite) between 0.001 to 2 V. Subsequently, dQ/dV analysis was performed based on this electrochemical data by taking the derivative of the capacity profile with respect to the voltage. Rate capability tests were performed under the same half-cell configuration by varying the rate every 5 cycles from C/20 to 2C and back to C/20. Commercial graphite and hard carbon

(BH-250) electrode powders were obtained from MTI Corp. and BTR New Energy Materials Inc., respectively.

### 2.5. Electronic conductivity measurements on pyrolyzed carbons

For each conductivity measurement, approximately 40-60 mg of carbon powder was packed into a ¼ inch diameter polyether ether ketone (PEEK) die using two stainless steel plungers. The die was secured into a cell holder and connected to a Solartron 1260 impedance analyzer under pressures of 0.10 and 0.15kN (around 3 and 5 MPa respectively). The average pellet size at 3 MPa was 2.46 ± 0.06 mm, and the average pellet size at 5 MPa was 2.21 ± 0.08 mm. The sample was held at 25°C for over an hour to allow for equilibration. Impedance measurements were then taken with an applied AC potential of 50 mV over a frequency range from 1 MHz to 1 Hz. The commercial carbon Super C65 used as reference was obtained from IMERYS (C-NERGY SUPER C65).

## 3. Results and Discussion

### 3.1. Structural Characterization using Raman, XPS, and ssNMR

The electrochemical properties of carbonaceous compounds are intimately linked to their purity, structural properties, and morphology (degree of graphitization, particle size, and surface area).[31] As shown in previous work, the structure and composition of MP carbon byproducts depend on the reaction conditions (temperature, catalyst, etc.).[4,7–14] The focus here is on identifying the links between carbon synthesis/processing, structure, and electrochemical behavior.

As summarized in **Table 1**, carbon byproducts formed during MP in the presence of catalytically-active $MnCl_2$-KCl and $FeCl_3$-NaCl-KCl salt mixtures are highly graphitic,[7,13] as evidenced by their low Raman D/G (or $I_D/I_G$) ratios. In contrast, carbons produced in the presence of less catalytically-active monovalent alkali halide salt environments (NaCl, KCl, NaBr, KBr) are significantly more disordered.[14,32,33]

| Carbon | Raman D/G ratio | Reference |
|---|---|---|
| NaCl, KCl, NaBr, KBr | 1 – 1.5 | [14,32,33] |
| FeCl$_3$ | 0.5 | [13] |
| MnCl$_2$ | 0.6 | [7] |

**Table 1.** Raman D/G ratios of carbons produced via methane pyrolysis using various molten salt catalysts.

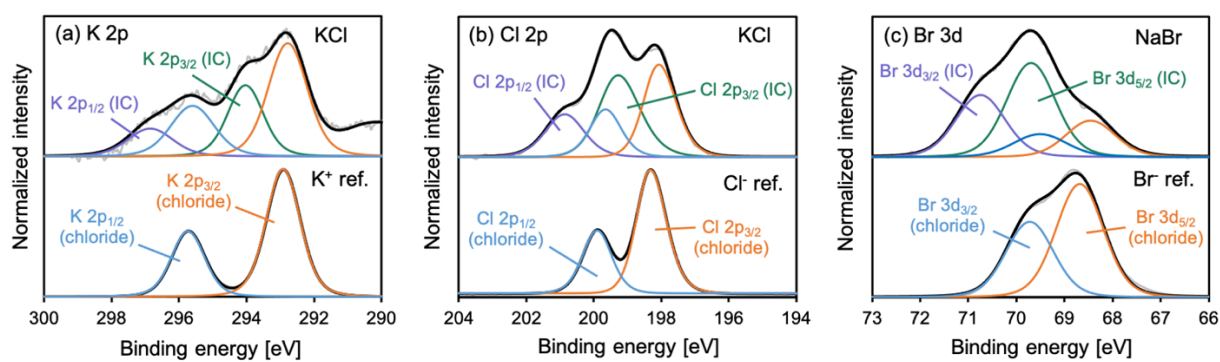

**Figure 1.** Spectroscopic identification of carbon-salt intercalation compounds (ICs) and salt units physically trapped into the carbon matrix. High-resolution XPS spectra of carbon samples synthesized from MP in molten salt bubble column reactors at 1000°C and reference spectra for alkali halide salts: (a) K 2p spectra of carbon synthesized in molten KCl (top) and of the reference KCl salt (bottom). (b) Cl 2p spectrum of carbon synthesized in molten KCl (top) and of the reference KCl salt (bottom). (c) Br 3d spectra of carbon synthesized in molten NaBr (top) and of the reference KBr salt (bottom).

In addition to the intrinsic physical characteristics of the carbon, potential residual salt in the carbon product is an important consideration for an economically viable commercial MP process and for the utilization/disposal of said carbon byproduct. Residual catalytic salts have been found to be difficult to remove from molten-salt-facilitated MP by post-processing (e.g., solvent extraction), which is presumably due to the inclusion of salt ions into the formed carbon interlayers.[7,13,14,33] While the difficulty in separating the carbon and salt post-reaction is detrimental to the economics of industrial MP, salt-retained carbon may be well-suited for electrochemical applications as-is. In fact, it is well known that graphite intercalation compounds (ICs) (such as KC$_8$ and LiC$_6$) exhibit good electrochemical properties, as well as more exotic properties such as superconductivity.[34]

To better understand carbon-salt interactions in MP byproducts, XPS measurements were conducted on the hot water washed KCl and NaBr carbons, and the latter system was further studied by ssNMR to gain complementary insights into its local structure. We selected these two representative systems as their constitutive elements are sensitive XPS (K, Cl, Br) and NMR ($^{13}$C, $^{23}$Na, and $^{71}$Br) probes. The potassium (K 2p) spectrum of the KCl carbon indicates the presence of potassium species in two distinct chemical bonding environments (**Figure 1a**). The major peak (K $2p_{3/2}$) of the characteristic 2p doublet at the lowest binding energy of ~292.7 eV in this spectrum (**Figure 1a**, top, orange curve) is close to that of a potassium ion (K$^+$) species in neat KCl salt (**Figure 1a**, bottom, orange curve). The major peak (K $2p_{3/2}$) of the second doublet, which is only present in the spectrum of the KCl carbon and not in that of the reference KCl salt, is centered at ~294.0 eV (**Figure 1a**, top, green curve). This new doublet has a marked upshift compared to the K$^+$ doublet and is in relatively good agreement with potassium intercalation compounds (ICs), such as KC$_8$ with a K $2p_{3/2}$ peak centered at 294.6 eV.[35] The slight discrepancy between the binding energy of this new doublet and that of the K $2p_{3/2}$ peak of KC$_8$ may suggest that the degree of graphitization and stoichiometry of the K-intercalated domains (i.e., ratio of K to C atoms) in the KCl carbon are lower than in KC$_8$, and/or arises from differences in the charge compensation processes. Regardless, the K 2p spectrum of KCl carbon suggest that K species are intercalated into the carbon network and that physically trapped KCl salt-like molecular units are also present. The Na 1s spectra obtained on the NaBr carbon and NaBr reference salt are not as informative, in part because the binding energy of photoelectrons emitted from core-like 1s orbitals of Na is less susceptible to differences in the chemical bonding environment than those from valence-like 2p orbitals of K, and because of the presence of sodium oxide contaminants in the reference salt despite our best efforts to prevent moisture exposure through sample handling in an inert environment before and during data acquisition. The Cl 2p spectrum collected on the KCl carbon (**Figure 1b**, top) and the Br 3d spectrum collected on the NaBr carbon (**Figure 1c**, top) are analogous to the K 2p spectrum obtained on the KCl carbon: one Cl (Br) doublet is similar to that observed for Cl$^-$ (Br$^-$) species in the KCl (KBr) reference salt (**Figure 1b,c**, bottom) and indicates salt-like molecular units physically trapped into the carbon matrix, and the other doublet is upshifted by ~1.2 eV. This second doublet is likely associated with intercalated Cl (Br) species, although no reliable reference spectra could be found in the literature. We can eliminate the possibility of Cl covalently bonded to C atoms giving rise to the upshifted doublet since organic chlorine compounds with such C-Cl bonds exhibit major Cl $2p_{3/2}$ signals centered at markedly higher binding energies (~200.5 eV)[36] than those observed in **Figure 1b**.

Overall, the XPS results strongly suggest that, besides physical trapping of salt-like molecular units, salt cations and anions are intercalated into the carbon matrix during MP. Yet, it is unclear from the XPS data alone whether the cations and anions intercalate separately or as a molecular unit. One might think that the molecular unit would be too bulky to be inserted into the structure, but large molecular ions, such as $FeCl_3^-$, are well-known graphite intercalants.[37] We therefore turn to ssNMR to shed further light on the local environments of carbon intercalants, focusing on the hot water washed NaBr carbon.

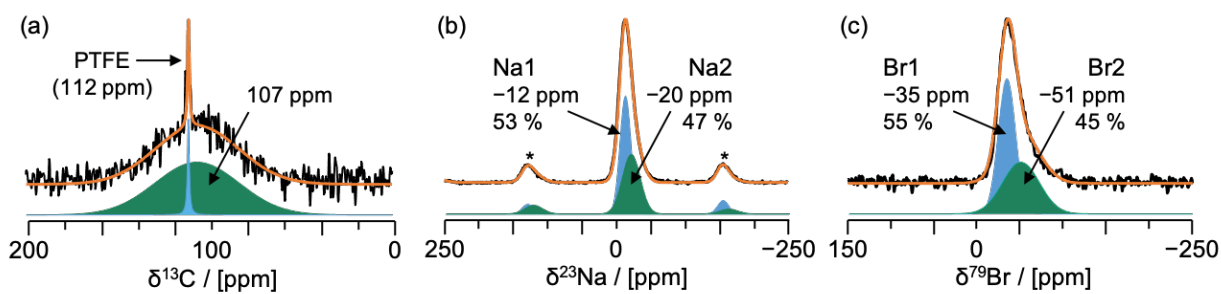

**Figure 2.** (a) $^{13}$C, (b) $^{23}$Na, and (c) $^{79}$Br spin echo ssNMR spectra collected on a hot water washed NaBr carbon sample. The ssNMR data were recorded under MAS (spinning speed $v_R$ = 30 kHz) and at $B_0$ = 18.8 T. Spinning sidebands are denoted with asterisks. For each spectrum, the overall fit is indicated with a yellow line, while individual signal components and their corresponding spinning sidebands are represented as green and blue shaded areas.

The $^{13}$C spin echo spectrum collected on the hot water washed NaBr carbon, shown in **Figure 2a**, allows us to directly probe the carbon structure. The spectrum can be deconvolved into two signals. The sharp signal at 112 ppm arises from the PTFE spacer used to center the carbon sample in the NMR rotor.[38] The second, very broad signal centered at 107 ppm is composed of many overlapping signals corresponding to a wide distribution of $sp^2$ and aromatic carbon environments[39–42] in the mostly disordered carbon (high D/G Raman signal intensity ratio in **Table 1**). These signals have a short lifetime ($T_2$ relaxation time), mostly due to the presence of many defects in the carbon structure that create a large number of spin relaxation/recombination pathways.

The intense $^{23}$Na ssNMR signal from carbon produced in molten NaBr indicates that a significant amount of Na is present in the sample even after a hot water wash (**Figure 2b**). The $^{23}$Na ssNMR spectrum exhibits a broad and asymmetric signal centered around -15 ppm, along with sidebands due to fast spinning of the sample during data acquisition. Notably, no sharp signal at 5 ppm could be observed, suggesting that no free NaBr salt is present in the sample.[43] The broad $^{23}$Na signal is composed of many overlapping $^{23}$Na resonances corresponding to a distribution of closely related Na environments in the sample, consistent with Na intercalation/trapping inside the disordered carbon. More specifically, the $^{23}$Na signal can be deconvolved into two broad Gaussian components (corresponding to two types of Na environments denoted Na1 and Na2) centered at −12 and −20 ppm and accounting for 53 and 47 % of the total integrated signal intensity, respectively. The spinning sidebands present in the $^{23}$Na NMR spectrum in **Figure 2** are indicative of residual chemical shift anisotropy (CSA) due to interactions with the carbon structure. Additionally, the observed negative $^{23}$Na shifts likely result from ring currents within graphene sheets or fragments that affect the local magnetic field experienced by nearby Na species.[44,45] Here, the closer the $^{23}$Na nucleus to the carbon framework, the more negative its chemical shift due to the additive nature of ring currents.[45] Hence, $^{23}$Na signals with a chemical shift around −12 ppm are tentatively attributed to salt-like Na$^+$ species trapped within the porous network of the carbon (Na1), while $^{23}$Na signals around −20 ppm are attributed to chemically intercalated Na$^+$ ions (Na2).

To better understand the origin of the broad $^{23}$Na spin echo signals in **Figure 2b**, additional and higher resolution $^{23}$Na spectra were recorded on the hot water washed NaBr carbon sample using the projected magic angle turning phase adjusted separation (pj-MATPASS)[25,26] and multiple quantum magic angle spinning (MQ-MAS)[27–30] experiments. While the pj-MATPASS experiment simplifies ssNMR spectra by suppressing spinning sidebands, the MQ-MAS experiment allows for the determination of the magnitude and distribution of $^{23}$Na quadrupolar coupling constants and isotropic chemical shifts associated with the various Na environments in the sample. Results from these additional experiments, shown in **Figure S1**, confirm that the broad signals centered around −12 and −20 ppm in the spin echo spectrum arise from a distribution of $^{23}$Na environments in the sample. Furthermore, the narrow range of $^{23}$Na quadrupolar parameters obtained from the MQ-MAS experiment suggests that all Na species are present in rather similar and low-symmetry environments (unlike Na in neat NaBr), consistent with

Na intercalated and trapped into disordered carbon.[46,47] More details on the analysis and interpretation of the pj-MATPASS and MQ-MAS data can be found in **Supplementary Note 1**.

The $^{79}$Br spin echo spectrum collected on the hot water washed NaBr carbon is shown in **Figure 2c** and exhibits a broad and asymmetric signal with no visible spinning sidebands. This signal spans more than 70 ppm and can be deconvolved into two Gaussian components (corresponding to two Br environments denoted Br1 and Br2) centered at −35 and −51 ppm, accounting for 55 and 45 % of the total integrated signal intensity, respectively. We now compare the distribution of Na and Br environments in the NaBr carbon sample, keeping in mind that the proposed deconvolution of the $^{23}$Na and $^{79}$Br spin echo spectra (in **Figure 2b** and **2c**, respectively) into two components is a highly simplified model for the actual broad distribution of Na/Br sites (as evidenced in **Figure S1** for Na). The ratio of Br1 to Br2 environments (55:45) is very close to that of Na1 to Na2 environments (53:47), and the distributions of Na2- and Br2-type environments (green) are broader than the distributions of Na1- and Br1-type environments (blue). It therefore appears that the distributions of Na and Br environments are correlated. A tentative interpretation of these results is that Na1- and Br1-type species are physically trapped together in the carbon framework, consistent with the salt-like molecular units observed by XPS, while Na2- and Br2-type species are co-intercalated into the carbon to satisfy local charge balance, leading to the upshifted XPS signals. Given the similar trends observed in the K, Cl, and Br XPS results discussed earlier, co-intercalated and physically trapped salt species are also found in the KCl carbon and presumably in all carbons produced by MP in molten salt media. Physical trapping of salt-like species takes place during formation of an extended carbon framework at > 1000°C. Previous work has shown that carbon micropores decrease in size with increasing formation temperature[48,49]. While carbon pores formed at temperatures above 1000°C still support Na$^+$ diffusion[49], and presumably K$^+$ diffusion (octahedral Na$^+$ and K$^+$ have ionic radii of 102 pm and 138 pm, respectively), they are likely too small to accommodate diffusion of the much larger Br$^-$ and Cl$^-$ species (with ionic radii of 196 and 181 pm, respectively).[50] Hence, we propose that the molten salt anions (Br$^-$, Cl$^-$) trapped inside newly-formed carbon cavities are unable to diffuse out and attract the smaller Na$^+$ and K$^+$ counterions that move through the micropores of the material during the pyrolytic process, creating salt-like molecular units in the bulk carbon pores. On the other hand, molten salt ions likely co-intercalate in-between graphene sheets or in curved graphene-like domains formed at > 1000°C, with a significant fraction of those domains becoming buried as the carbon framework grows. It then becomes obvious that a hot water wash is rather ineffective at cleaning the carbons of interest as it can only remove surface ions and is unable to extract most of the salt species in the bulk of the hydrophobic carbon.

Overall, direct NMR characterization of the carbon materials of interest is challenging due to the intrinsically low sensitivity of natural abundance $^{13}$C NMR conducted on non-protonated materials.[41,51] This issue is circumvented here through the use of multi-nuclear NMR spectroscopy, in combination with XPS, providing insights into the environments of NMR-active (e.g., $^{23}$Na and $^{79}$Br) catalytic molten salt species in pyrolyzed carbon. Interesting directions for future work include probing the interactions between the carbon structure (after $^{13}$C enrichment) and the intercalated/trapped salt ions/units in greater detail, and optimizing the pyrolysis conditions (e.g., temperature profile) to facilitate removal of residual salt from the structure.

### 3.2. Electrochemical Properties of Pyrolyzed Carbon Electrodes

For applications in alkali-ion batteries, carbon electrodes must exhibit a large and reversible charge storage capacity, a low working potential, and a good capacity retention over extended electrochemical (charge-discharge) cycling. In other words, the carbon structure must be amenable to reversible bulk intercalation of alkali ions at low potentials. To assess the electrochemical properties of the pyrolyzed carbons of interest to this work, galvanostatic cycling tests were combined with differential capacity (dQ/dV) analyses and rate capability tests.

We first determined the impact of hot water washing and of an additional washing step using $NO_2BF_4$ in a 0.1 M acetonitrile solution on electrochemical Li insertion into / extraction from carbons obtained from MP through galvanostatic cycling of Li-ion half cells. The additional $NO_2BF_4$ washing step was employed here as it has been found to be successful in extracting ions from intercalation compounds. [52–56] The initial discharge corresponds to Li stripping at the Li metal counter-electrode (oxidation process) and insertion into the carbon electrode (reduction process), while opposite processes take place on subsequent charge. The first cycle electrochemical profiles, reversible capacities, and average working potentials of the pristine and washed carbons are compared in **Figures S2 and S3**, with values listed in **Tables S1-2**. The first cycle reversible capacities do not evolve significantly nor consistently with washing. Similarly, for most carbons, no clear trend could be established between the extent of washing and small (≤

0.1 V) changes in the Li (de)intercalation potential. These small and inconsistent changes in the electrochemical properties may be accounted for by the presence of several contributions to a carbon electrode's total capacity and average potential that evolve in opposite directions upon washing, and because our washing procedures are unable to fully remove molten salt residues from the carbon matrix, as discussed earlier. Yet, we find that after each additional washing step the voltage profiles become better defined, suggesting that the washing steps do remove some amount of salt impurities from the carbon structure. Notably, we find that residual redox-active catalytic salts in the $FeCl_3$ and $MnCl_2$ carbons undergo a conversion-type electrochemical reaction on initial discharge, a process that is only partially reversible on subsequent charge and consumes a significant amount of Li (see **Figure S4**). These processes become more reversible and/or less prevalent after a hot water wash, likely due to partial salt oxidation and removal from the carbon. A more detailed analysis of the impact of washing on the electrochemistry of the pyrolyzed carbons, and of the conversion reactions involving $FeCl_3$ and $MnCl_2$ salts, can be found in **Supporting Note 2**. Overall, our results suggest that a hot water wash of the pyrolyzed carbons is beneficial to reduce irreversible electrochemical reactions consuming Li. The additional $NO_2BF_4$ washing step only seems to remove a small amount of $MnCl_2$ salt from the carbon and does not improve the performance of the other carbons. Hence, washing with $NO_2BF_4$ is not cost-effective for large-scale applications and we hereafter focus on the electrochemical behavior of the hot water washed carbons.

The reversible Li-ion intercalation capacities and average working potentials of the hot water washed pyrolyzed carbons are shown in **Figure 3** along with values obtained for commercial graphite and hard (disordered) carbon electrodes. The reversible capacity of the carbons depends on their degree of graphitization and purity, and increases as follows: NaBr C < KCl C < NaCl C < $FeCl_3$ C < $MnCl_2$ C < hard C < graphite. The corresponding electrochemical curves (plots of the voltage vs. specific capacity) and differential capacity (dQ/dV) plots are shown in **Figure S5**. Peaks in the dQ/dV plots correspond to discrete electrochemical processes and provide a sensitive picture of the Li-ion (de)intercalation mechanism. All carbons, including commercial graphite and hard carbon, exhibit a large irreversible capacity during the first discharge-charge cycle. This irreversible capacity can be at least partly attributed to the formation of a solid electrolyte interphase (SEI) upon decomposition of the electrolyte salt ($LiPF_6$) or of the ethylene carbonate/dimethyl carbonate (EC/DMC) solvent mixture at the surface of the carbon on initial discharge.[57,58] These side reactions are associated with negative dQ/dV peaks in the range of 0.9 to 0.2 V vs. Li$^+$/Li for commercial graphite and the FeCl$_3$ and MnCl$_2$ carbons, and in the range of 1.2 to 0.4 V for hard carbon and the NaBr, NaCl, and KCl carbons. The dQ/dV peaks are more intense for the pyrolyzed carbons than for the commercial graphite and hard carbon electrodes, suggesting that residual molten salts at the surface of or intercalated/trapped inside the carbon catalyze electrolyte decomposition reactions.[59] Hence, a thicker SEI layer is expected to form at the surface of the pyrolyzed carbons, likely contributing to the observed irreversible capacity.

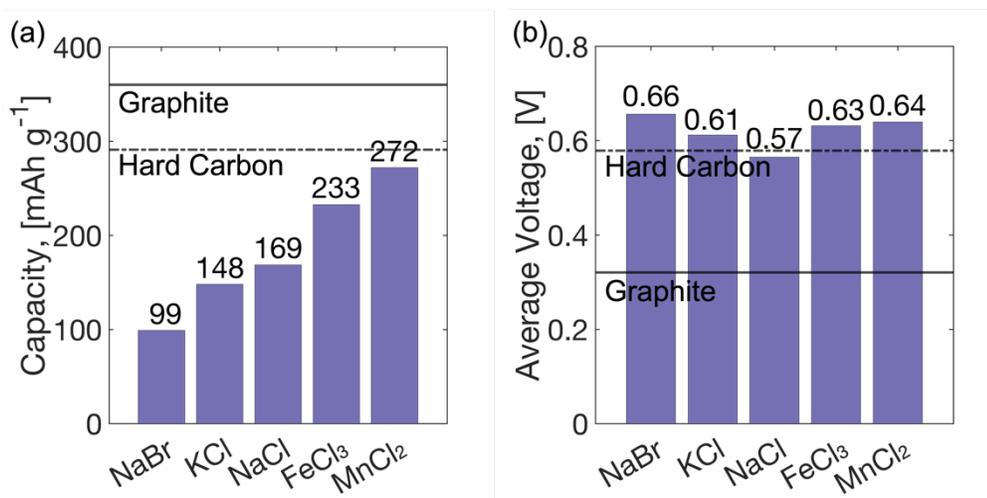

**Figure 3.** Electrochemical properties of hot water washed carbons obtained via MP using various molten salt catalysts. (a) First discharge-charge cycle reversible capacities and (b) average Li extraction potentials of the carbon electrodes tested in Li-ion half cells. The performance of commercial graphitic and hard carbon electrodes is shown with solid and dashed lines, respectively.

The highest reversible capacity and lowest average Li deintercalation potential are obtained for the MnCl$_2$ carbon, followed by the FeCl$_3$ carbon. As shown in **Figure S5**, the MnCl$_2$ carbon can deliver a reversible capacity up to 75% of that of graphite (360 mAh/g) and up to 96% of that of hard carbon (290 mAh/g) when cycled at a rate of C/20 (based on graphite) between 0.001 and 2 V vs. Li$^+$/Li. Its working potential is on par with that of hard carbon but remains 0.25 V higher than that of graphite, presumably due to the presence of both disordered and graphitic domains in the structure, as well as a small amount of electrochemically active MnCl$_2$ salt.

Given the high sensitivity of the Li-ion (de)intercalation potential on the degree of graphitization of carbon electrodes, the electrochemical response of pyrolyzed carbons can be used as a qualitative tool to determine the degree of ordering of component graphene sheets. The two structural end-members are graphite, composed of an ordered stacking of graphene layers, and hard carbon, consisting of a random arrangement of curved graphene sheets. The pyrolyzed carbons fall into two broad categories based on their dQ/dV profiles: 'hard carbon-like' for the NaBr, NaCl, and KCl carbons, and 'graphitic' for the $MnCl_2$ and $FeCl_3$ carbons. These findings are in line with the Raman results presented in **Table 1**. The dQ/dV plots obtained during the second Li extraction process (charge) for the NaBr, KCl, and NaCl carbons are overlaid with that of hard carbon in **Figure 4**. All dQ/dV curves exhibit a single peak and look alike, suggesting that Li storage occurs via an intercalation-adsorption or via an adsorption-intercalation mechanism, as has been suggested for hard carbon.[60] Differences in peak intensities are related to the different Li storage capacities of the carbons under study, while differences in Li insertion potentials are likely due to slight differences in electrode loading density.[61] The results shown in **Figure 4** indicate that Li intercalation into graphitic domains is relatively low for the NaBr, KCl, and NaCl carbons, as evidenced by their small dQ/dV peaks below 0.2 V vs. $Li/Li^+$. Given the disordered nature of the carbons discussed here and the propensity of such carbons to intercalate $Na^+$ ions,[62–64] the KCl carbon was also tested in a Na-ion half-cell (see **Figure S6**). However, a reversible capacity of 30 mAh/g, much lower than that of a commercial hard carbon electrode (226 mAh/g), was obtained.

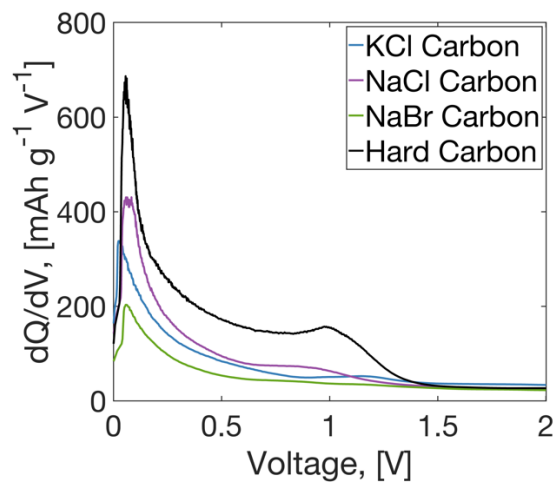

**Figure 4.** dQ/dV plots for hard carbon, and KCl, NaCl, and NaBr carbons during the second Li extraction process (charge) at a rate of C/20 (based on a graphite electrode) and over a voltage window from 0.001 V to 2V vs. $Li/Li^+$.

The dQ/dV plots obtained during the second charge process for the $MnCl_2$ and $FeCl_3$ carbons are overlaid with that of graphite in **Figure 5a**. Three major dQ/dV peaks are observed in the 0.25-0.005 V range for the reference graphite anode, and to a lesser extent in the dQ/dV curves of the $MnCl_2$ and $FeCl_3$ carbons. It has been shown that Li-ion (de)intercalation into graphite occurs in sequential stages first identified by Ohzuku et al.[65] Four Li-ion intercalation regions corresponding to transitions between stages 1, 2, 3, 4, and 8, are summarized below:

Region I (85 mV): $LiC_6$ ($1^{st}$ stage) = $LiC_{12}$ ($2^{nd}$ stage)
Region II (120 mV): $LiC_{12}$ ($2^{nd}$ stage) = $LiC_{18}$ ($2^{nd}$ stage)
Region III (120-210 mV): $LiC_{18}$ ($2^{nd}$ stage) = $LiC_{27}$ ($3^{rd}$ state) = $LiC_{36}$ ($4^{th}$ state)
Region IV (210 mV): $LiC_{36}$ ($4^{th}$ stage) = $LiC_{72}$ ($8^{th}$ stage)

The three dQ/dV peaks in are attributed to regions I, II, and IV, as annotated in **Figure 5a**, and the observed variations in the dQ/dV peak positions are once again attributed to differences in electrode loading densities.[61] The low intensity of the dQ/dV peaks in the data collected on the $MnCl_2$ and $FeCl_3$ carbons suggests that these carbons are not fully graphitic and contain some disordered domains. In **Figure 5b**, we thus compare the dQ/dV curves obtained for the $MnCl_2$ and $FeCl_3$ carbons to that obtained for hard carbon during the second charge process. Below 0.05V, all three carbons show a broad dQ/dV feature that is absent from the dQ/dV curve of pure graphite (**Figure 3a**). This broad feature is more intense in the case of hard carbon and is indicative of disordered domains present in the $MnCl_2$ and $FeCl_3$ carbons. After subtracting the capacity from side reactions, at least 242 mAh/g of capacity can be attributed to Li-ion (de)intercalation from/into the $MnCl_2$ carbon structure, and predominantly into graphitic domains. Similarly, at least 220 mAh/g of the capacity is coming from graphitic domains in the $FeCl_3$ carbon structure.

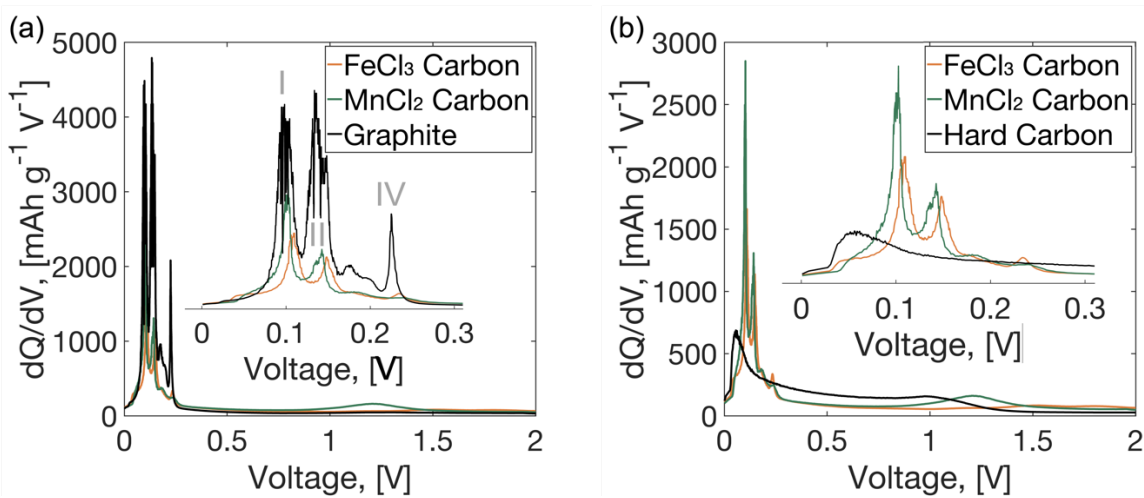

**Figure 5.** dQ/dV plots of different carbons during the second Li extraction process (charge) at a rate of C/20 (based on a graphite electrode) and over a voltage window from 0.005V to 2V vs. Li/Li$^+$. **a)** Comparison of graphite with FeCl$_3$ and MnCl$_2$ carbons. **b)** Comparison of hard carbon with FeCl$_3$ and MnCl$_2$ carbons.

The rate capability of carbon-based Li-ion electrodes, or their ability to rapidly (de)intercalate Li-ions, is a critical parameter for high power applications. Here, MnCl$_2$ and KCl carbons were selected for rate capability testing due to their contrasting structural properties, and their specific capacity and capacity retention at various current rates are compared to those of commercial graphite in **Figure 6**. The graphite electrode performs significantly better than the pyrolyzed carbons at ≤ 1C rates (**Figure 6a**). At current rates slower than C/5, all carbons show good capacity retention (~90%+) although graphite performs the best (**Figure 6b**). As the current rate is increased to 1C, the capacity retention drops significantly for all carbons, with the KCl carbon exhibiting the best capacity retention, presumably due to its low overall electrode use (capacity < 100 mAh/g). It is important to note that the low rate capability of the cells tested here may in part be due to the use of a commercial electrolyte (LP30) with a low Li$^+$ transference number, resulting in large concentration gradients and sluggish kinetics at rates above 1C.[66,67] To better understand the origin of the low-rate capability of the pyrolyzed carbons, dQ/dV plots obtained for the MnCl$_2$ carbon at current rates of C/20 and 1C are shown in **Figure S7a**. Peaks corresponding to Li intercalation into the graphitic and amorphous domains of the structure almost disappear at 1C, indicating sluggish Li-ion transport in all regions of the carbon. Next, we tested the impact of an additional NO$_2$BF$_4$ wash of the MnCl$_2$ and KCl carbons on their rate performance, with results shown in **Figures S7b,c.** The additional washing step improves the rate capability of both carbons at rates of 1C and above, suggesting that residual catalytic salts at the surface of the pyrolyzed carbons significantly affect Li-ion transport in and out of the electrode particles.

Our investigations of the electrochemical properties of pyrolyzed carbon anodes indicate that the carbon produced in a catalytic molten salt mixture containing $MnCl_2$ is potentially useful in Li-ion battery anode applications, especially for lower power applications such as in low-cost grid storage. This carbon was observed to have reasonable capacity and rate performance, and good capacity retention. Given that all electrochemical performance metrics considered here are negatively impacted by the presence of salt residues, the ability to obtain cleaner, higher value-added carbons will be an important step towards making MP economically competitive with current methods of hydrogen production.

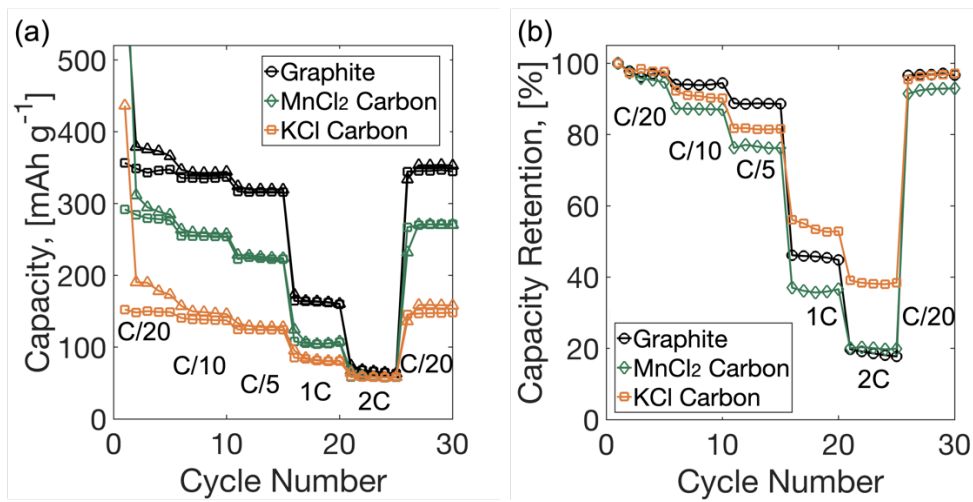

**Figure 6.** Rate capability of selected carbons over a voltage window from 0.001 V to 2V vs. Li/Li$^+$. **a)** Specific capacity, and **b)** capacity retention, observed for KCl carbon, $MnCl_2$ carbon, and graphite upon step increases in current rate from C/20 to 2C and back to C/20 over 30 cycles.

### 3.3. Pyrolyzed Carbons as Conductive Additives

While graphitic carbons are optimal for use as Li-ion electrodes due to their high capacities at low potentials, so-called carbon blacks (e.g., Super C65 and Super P) composed of a mixture of graphitic and disordered domains[68] provide higher electronic conductivities than pure graphite and are used as conductive additives in composite electrodes.[69–72] The demand for carbonaceous conductive additives, which make up 5-20 wt.% of composite electrodes, is expected to soar concurrently with that of Li-ion batteries,[22] providing another potential high value application for carbons produced via MP. We thus evaluated the electronic conductivity of the hot water washed carbons using electrochemical impedance spectroscopy (EIS).

The electronic conductivities of the various carbons of interest are reported at two different cell pressures of 3 and 5 MPa and compared to those obtained for a commercial conductive additive (Super C65) in **Table 2**. The electronic conductivity of the carbon samples increases as follows: KCl ~ $MnCl_2$ ~ $FeCl_3$ < NaBr < NaCl < Super C65, where the conductivity of the NaCl carbon is about 80-85% of that of carbon Super C65. There is no clear trend between the electronic conductivity and degree of graphitization. Instead, the single most predictive factor for conductivity is carbon purity. This is shown in **Table 2** where the purity of the pyrolyzed carbons obtained in previous studies using energy dispersive X-ray (EDX) analysis, thermogravimetric analysis (TGA), and/or X-ray fluorescence (XRF), is reported.[7,13,33] While the presence of mostly insulating salt impurities likely contributes to the lower electronic conductivity of the pyrolyzed carbons as compared to carbon Super C65, the surface area of conductive carbon additives has also been shown to play a key role in their performance, and differences in the average particle size of the carbons considered here will affect their compressibility and measured conductivities. BET measurements published on similar pyrolyzed carbons[33] indicate that their surface area (~2-3 $m^2/g$) is vastly lower than that of the commercial carbon Super C65 used here (62 $m^2/g$). We note, however, that such measurements are affected by the salt contaminants in terms of both sample weight and surface area. Scanning electron microscopy (SEM) images reported elsewhere[14] suggest that the pyrolyzed carbons have a particle size of at least ~500 nm (compared to < 50 nm for carbon Super C65) but a higher surface area than the values obtained from the BET measurements, complicating our interpretation of the observed electronic conductivity trend. Yet, with the results presented here, we once again conclude that developing better synthetic strategies to prevent salt incorporation into the carbon matrix at high temperatures is key to the valorization of the carbon byproduct of MP and its implementation for hydrogen production.

| Carbon | MnCl$_2$* | FeCl$_3$ | KCl | NaBr | NaCl | C65 |
|---|---|---|---|---|---|---|
| Conductivity @ 3 MPa (S/cm) | 0.107 | 0.111 | 0.109 | 0.124 | 0.229 | 0.285 |
| Conductivity @ 5 MPa (S/cm) | 0.110 | 0.148 | 0.120 | 0.157 | 0.284 | 0.332 |
| Purity (Atomic %) | EDX: 96.22 ± 0.76 [7]<br>XRF: 93.15 ± 0.82 [7] | EDX: 92.83 ± 2.61 [13] | EDX: 91.97 ± 2.41 [7]<br>XRF: 92.39 ± 1.03 [13]<br>TGA: 92.2 [33] | TGA: 96.1 [33] | TGA: 97.8 [33] | N/A |

*Results obtained on carbon pyrolyzed using a 2:1 ratio of KCl:MnCl$_2$ salt, while the carbon studied here was obtained from a 1:1 KCl-MnCl$_2$ molten salt mixture.

**Table 2.** Electronic conductivities of pyrolyzed carbons and of a commercial carbon additive reference (carbon Super C65) at 3 and 5 MPa. These conductivities are compared against the purity of the carbons quantified using EDX, TGA, or XRF.

## 4. Conclusions

Methane pyrolysis (MP) can be used to produce hydrogen and solid carbon without direct carbon dioxide emissions. Yet, MP can only compete with existing, greenhouse gas-emitting, methane reforming processes if the carbon co-product can be valorized, and/or by imposing a carbon tax. In this work, we assessed the applicability of carbons obtained from MP as high value-added anodes or conductive additives for secondary Li-ion and Na-ion batteries. XPS and solid-state NMR characterization of carbons obtained using various molten salt catalysts provided unprecedented detail on their short-range structures and revealed the presence of physically-trapped salt-like units and co-intercalated salt cations and anions in the carbon matrix, even after aggressive washing procedures. Consistent with previous work, Raman spectroscopy and differential capacity analysis in Li-ion and Na-ion half cells indicated that more graphitic pyrolyzed carbons formed when using catalytically-active FeCl$_3$– and MnCl$_2$–containing salts. These two carbons also exhibited the highest capacities in Li-ion half cells (272 and 233 mAh/g, respectively). While promising, these reversible capacities remain lower than that of a reference graphite Li-ion anode, partially because all pyrolyzed carbons contain disordered domains, but mostly due to residual catalytic salt impurities from the pyrolysis process. The presence of residual salt in the pyrolyzed carbons also results in higher Li-ion cell operating potentials (~0.57-0.66 V vs. $Li^+/Li$) compared to graphite (~0.32 V vs. $Li^+/Li$), thus reducing their energy density, to more rapid capacity decay upon charge-discharge cycling, and to a poorer rate performance at ≥ 1C. Further, the electronic conductivity of the pyrolyzed carbons depends sensitively on their purity, with the higher purity (NaCl) carbon exhibiting a conductivity of about 80-85% of that of standard carbon conductive additives. While there is a clear potential to valorize MP through the deployment of the carbon byproducts in secondary batteries, future work should focus on identifying ways to either fully remove salt impurities post-pyrolysis, better separate the carbon from the molten salt at high temperatures, or find alternative catalysts for MP with a lower tendency to mix with the carbon.

## Acknowledgements

This work was primarily supported by the Office of Energy Efficiency and Renewable Energy (EERE), Fuel Cell Technologies Office, of the U.S. Department of Energy, under contract No. DE-EE0008845. The NMR and XPS results reported here made use of shared facilities of the UCSB MRSEC (NSF DMR # 1720256), a member of the Materials Research Facilities Network (www.mfn.org). E. E. Foley was supported by an NSF Graduate Research Fellowship under Grant No. DGE 1650114. C. Palmer was supported by the Energy & Biosciences Institute through the EBI-Shell program. The authors are grateful to Ms. Ashlea Patterson for her assistance with XPS data collection.

## Supporting Information Available

Additional characterization of hot water washed carbons using solid-state NMR; Electrochemical Characterization of Pyrolyzed Carbons.

# Valorizing the carbon byproduct of methane pyrolysis in batteries

## Supplementary Information

Ji, Y.,[a,b] Palmer, C.,[c] Foley, E. E.,[a,b] Giovine, R.,[a,b] Yoshida, E.,[a,b]
McFarland, E.,[c] and Clément, R. J.[a,b*]

a. *Materials Department, University of California Santa Barbara, California 93106, United States*
b. *Materials Research Laboratory, University of California Santa Barbara, California 93106, United States*
c. *Chemical Engineering Department, University of California Santa Barbara, California 93106, United States*

*Corresponding author. Tel: 805-893-4294. E-mail: rclement@ucsb.edu (Raphaële J. Clément)

# Additional characterization of hot water washed carbons using solid-state NMR

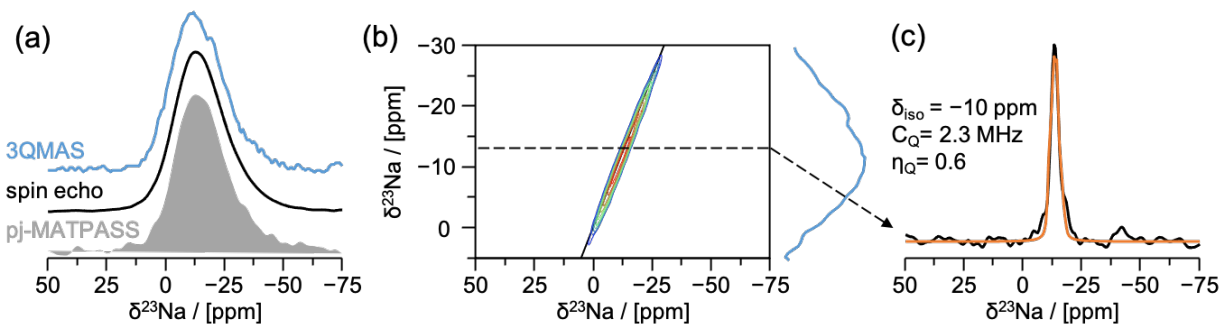

**Figure S1.** $^{23}$Na ssNMR spectra collected on the hot water washed NaBr carbon sample. (a) From top to bottom: 3QMAS anisotropic projection (light blue line obtained from the 2D 3QMAS data presented in (b)), spin echo spectrum (black line), and fully isotropic pj-MATPASS spectrum (gray shaded area)[1,2]. (b) High resolution $^{23}$Na z-filter 3QMAS sheared 2D spectrum [3–6]. The projection of the isotropic dimension is shown in light blue to the right of the spectrum and in (a). (c) 1D slice extracted from the 3QMAS 2D spectrum at the chemical shift position indicated with a black dashed line in (b). The experimental spectrum is shown in black and its fit in yellow. All NMR spectra were recorded at $B_0$ = 18.8 T and at a MAS speed $\nu_R$ = 30 kHz.

## Supplementary Note 1: $^{23}$Na pj-MATPASS and MQ-MAS results.

To better understand the origin of the broad and overlapping $^{23}$Na spin echo signals (denoted as Na1 and Na2) in **Figure 2a**, additional and higher resolution $^{23}$Na spectra were recorded on the hot water washed NaBr carbon sample using the projected magic angle turning phase adjusted separation (pj-MATPASS)[1,6] and multiple quantum magic angle spinning (MQ-MAS)[3–6] experiments. First, an enlarged view of the central region of the $^{23}$Na spin echo spectrum (middle spectrum in **Figure S1a**, in black) is compared to the fully isotopic spectrum obtained from pj-MATPASS (bottom spectrum in **Figure S1a**, represented as a gray shaded area). The similar spectral lineshapes and linewidths obtained with these two experiments indicate that the line broadening observed in the spin echo spectrum does not result from sideband overlap, but instead arises from a distribution of closely related Na environments, leading to many overlapping signals with similar resonant frequencies, and/or from quadrupolar interactions between $^{23}$Na nuclei ($I$ = 3/2) and electric field gradients created by nearby charged species. Contributions from these two potential sources of linebroadening are separated in a triple quantum excitation (3QMAS, a variant of MQMAS) ssNMR experiment discussed hereafter. The 1D anisotropic projection of the 2D 3QMAS spectrum collected on the NaBr carbon sample is shown in **Figure S1a** (top spectrum in light blue) and its lineshape is similar to the spin echo and pj-MATPASS spectra, confirming that all $^{23}$Na signals are efficiently excited in the 3QMAS experiment. The 2D $^{23}$Na 3QMAS spectrum is shown in **Figure S1b** and exhibits an elongated signal along the diagonal that is characteristic of a distribution of isotropic $^{23}$Na chemical shifts, confirming the presence of multiple Na environments in the sample. The 3QMAS data also indicate that the quadrupolar parameters (quadrupolar coupling constant ($C_Q$) and asymmetry ($\eta_Q$)) of all of the $^{23}$Na signals fall within a narrow range of values, suggesting that all Na environments are very similar. A fit of a 1D spectrum obtained from a horizontal slice through the 3QMAS 2D spectrum (see **Figure S1c**) leads to a rather large $C_Q$ value (2.3 MHz) and $\eta_Q$ (0.6), corresponding to a single, low symmetry (unlike Na in neat NaBr) Na site that is consistent with Na intercalated and/or trapped in the carbon structure.[7,8]

**Electrochemical Characterization of Pyrolyzed Carbons**

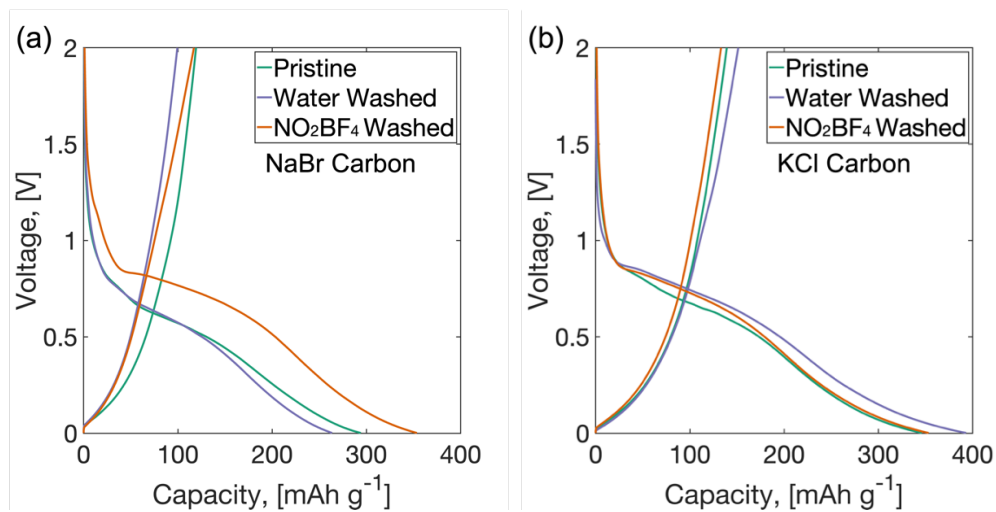

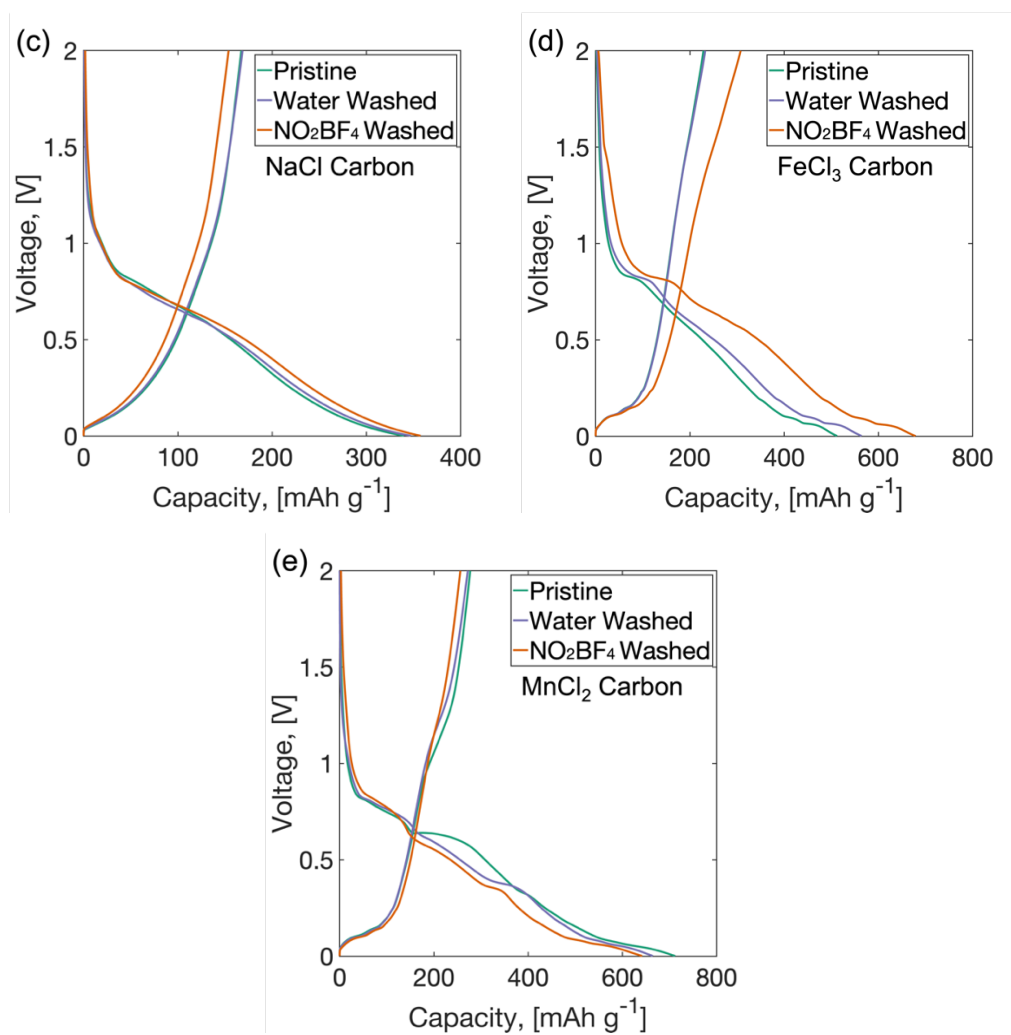

**Figure S2.** Comparison of the Li-ion half-cell first cycle voltage vs. capacity curves of pyrolyzed carbon electrodes in the pristine state (green), after a hot water wash (purple), and after a hot water + $NO_2BF_4$ wash (orange). a) NaBr carbon, b) KCl carbon, c) NaCl carbon, d) $FeCl_3$ carbon, and e) $MnCl_2$ carbon. All cells were cycled over the 0.001-2 V vs. $Li^+/Li$ potential range at a rate of C/20 (based on graphite).

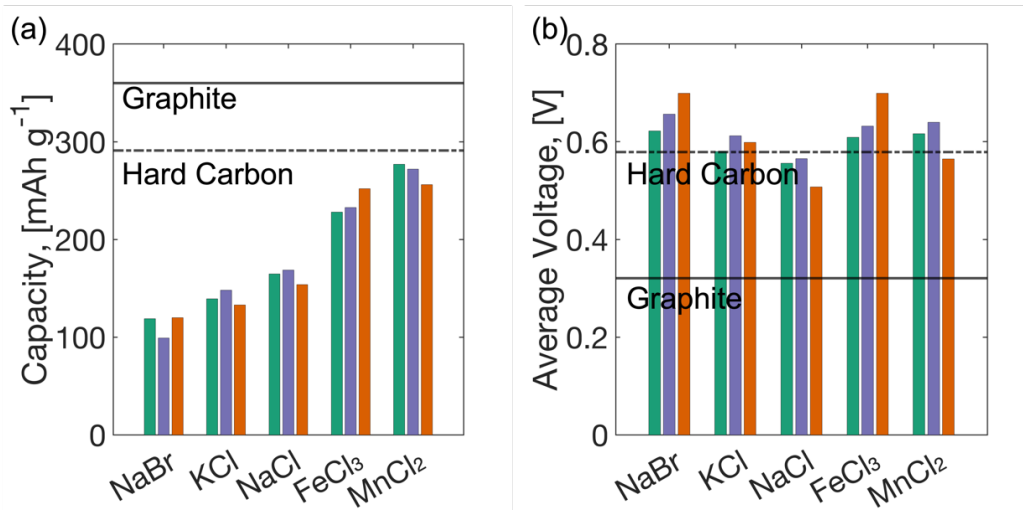

**Figure S3.** (a) First cycle reversible capacity and (b) average lithium extraction potential obtained in Li-ion half cells for pyrolyzed carbon electrodes in the pristine state (green), after a hot water wash (purple), and after a hot water + $NO_2BF_4$ wash (orange). The performance of reference hard carbon and graphite electrodes is shown with dashed and solid lines, respectively. All cells were cycled over the 0.001-2 V vs. $Li^+/Li$ potential range at a rate of C/20 (based on graphite electrode).

|  | Pristine [mAh/g] | Water washed [mAh/g] | Water + $NO_2BF_4$ washed [mAh/g] |
|---|---|---|---|
| NaBr Carbon | 119 | 99 | 120 |
| KCl Carbon | 139 | 148 | 133 |
| NaCl Carbon | 165 | 169 | 154 |
| $FeCl_3$ Carbon | 228 | 233 | 252 |
| $MnCl_2$ Carbon | 284 | 285 | 256 |

**Table S1.** Summary of the reversible capacity of pyrolyzed carbons at different stages of washing, corresponding to the results shown in **Figure S2**.

|  | Pristine [V] | Water washed [V] | Water + $NO_2BF_4$ washed [V] |
|---|---|---|---|
| NaBr Carbon | 0.6223 | 0.6566 | 0.6991 |
| KCl Carbon | 0.5797 | 0.6122 | 0.5986 |
| NaCl Carbon | 0.5560 | 0.5652 | 0.5077 |
| $FeCl_3$ Carbon | 0.6089 | 0.6318 | 0.6991 |
| $MnCl_2$ Carbon | 0.6162 | 0.6597 | 0.5651 |

**Table S2.** Summary of the average Li extraction potentials of pyrolyzed carbons at different stages of washing, corresponding to the results shown in **Figure S2**.

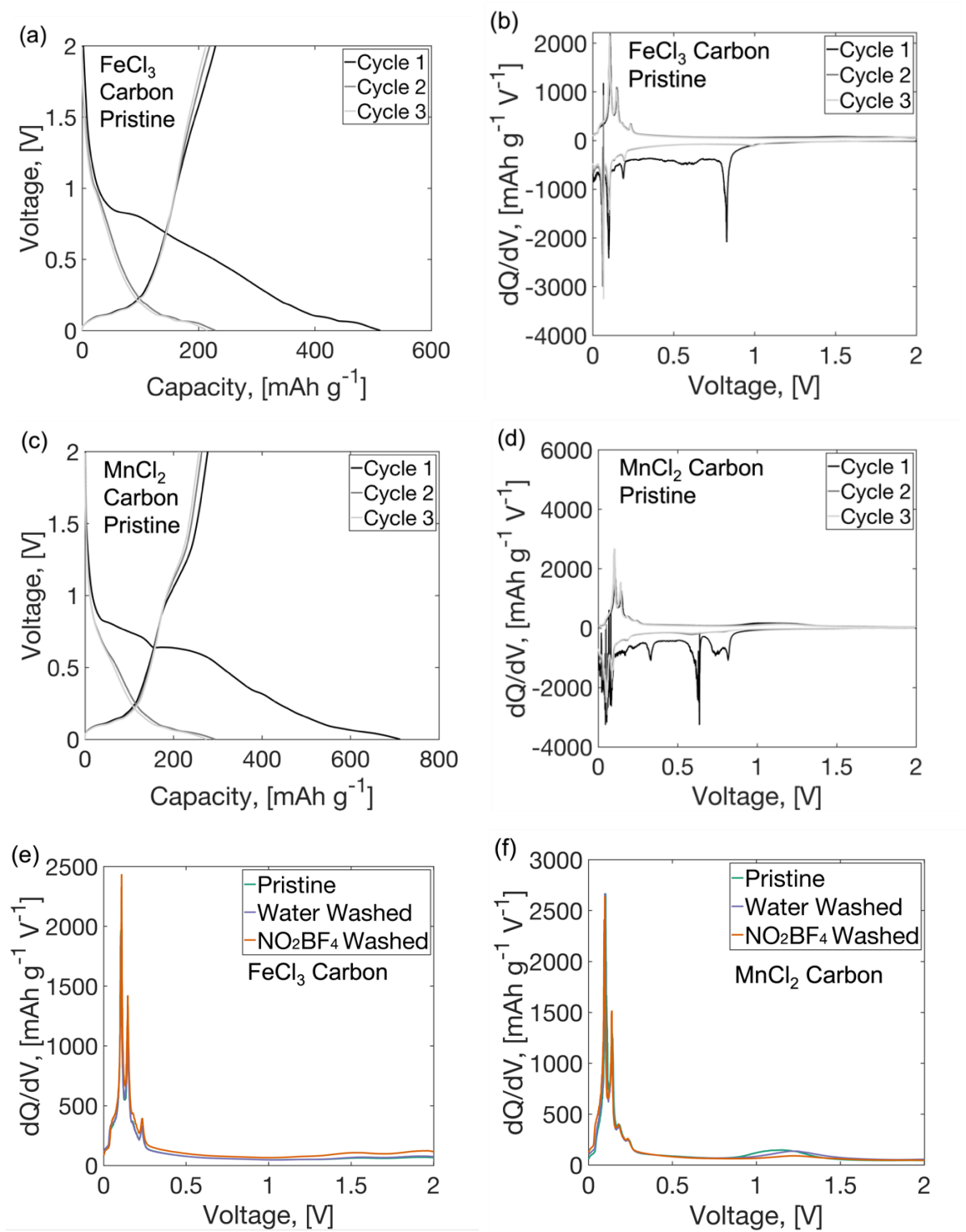

**Figure S4.** Evolution of the electrochemical properties of FeCl$_3$ and MnCl$_2$ carbon electrodes with washing. Li-ion half cell voltage profiles for the first 3 discharge-charge cycles obtained on pristine (a) FeCl$_3$ and (c) MnCl$_2$ carbon electrodes. Their differential capacity (dQ/dV) plots are shown in (b) and (d), respectively. Comparison of the dQ/dV curves obtained upon charge of the (e) FeCl$_3$ and (f) MnCl$_2$

carbon electrodes in the pristine state (green), after a hot water wash (purple), and after a hot water + NO$_2$BF$_4$ wash (orange). Electrochemical tests were carried out over a voltage window from 0.001 V to 2 V vs. Li/Li$^+$ and at a current rate of C/20 (based on graphite).

**Supporting Note 2: Impact of washing on the electrochemical performance of carbons obtained via MP.**

A two-step process was devised to remove residual catalytic salt impurities from the carbons obtained via MP, and the effect of each washing step on the electrochemical performance of the carbons was investigated. First, the pristine carbons were washed with 70°C water for 2 hr. In a second step, the carbons were immersed in a 0.1 M NO$_2$BF$_4$ in acetonitrile solution with constant stirring for 24 hr to further remove intercalated or adsorbed salts, as this procedure has been found to be successful in extracting charged species from related intercalation compounds.[9–13]

The electrochemical performance of the pristine, water washed and water + NO$_2$BF$_4$ washed carbons were tested in Li-ion half-cells using galvanostatic cycling. The first cycle electrochemical profiles for each carbon at different stages of washing are shown in **Figure S2**. The first cycle reversible capacities and average Li extraction potentials of the pristine and washed carbons are compared in **Figure S3**, with exact values listed in **Tables S1-2**. The first cycle reversible capacities do not evolve significantly nor consistently upon washing. Yet, after each additional washing step, the voltage profiles become better defined, suggesting that the washing may remove some amount of salt impurities from the carbon structure. The average working potentials evolve by up to 0.1 V upon washing, but for most carbons, no clear trend can be established between the extent of washing and the changes in Li (de)intercalation potential. We attribute these small and inconsistent changes to the presence of several contributions to a carbon electrode's total capacity and average potential that evolve in opposite directions upon washing. While the removal of residual catalytic salt is expected to enhance the reversible capacity of the NaCl, KCl, and NaBr carbons by rendering their pores more accessible to Li$^+$ ion intercalation, we find instead that the weakly bound surface impurities that can be removed through washing do not have a significant effect on the reversibility of the electrochemical processes. On the other hand, the reversible capacity of the FeCl$_3$ and MnCl$_2$ carbons could either increase or decrease upon washing, depending on whether the dominant capacity change arises from graphitic/disordered carbon domains becoming more accessible for Li (de)intercalation, due to fewer surface salt residues and a thinner SEI, or from reduced conversion-type electrochemical reactions involving the electrochemically-active $FeCl_3$/$MnCl_2$ salts. The latter electrochemical processes can readily be observed in the electrochemical curves and dQ/dV plots obtained during the first three cycles for the pristine $FeCl_3$ / $MnCl_2$ carbons, shown in **Figure S4**. For the $FeCl_3$ carbon, a voltage plateau (**Figure S4a**) and a sharp dQ/dV feature (**Figure S4b**) are observed between 0.9 and 0.8 V vs. Li/Li$^+$ on initial discharge, and closer to 1 V vs. Li/Li$^+$ on subsequent discharges. This evolution of the electrochemical features in the 0.8-1 V vs. Li/Li$^+$ region is reminiscent of the $FeCl_3$ + 3Li$^+$ + 3e$^-$ → Fe + 3LiCl conversion process reported for an $FeCl_3$-graphite intercalation compound.[21-22] Hence, we attribute these features to the conversion of $FeCl_3$ into LiCl and Fe in domains of $FeCl_3$-intercalated graphite within the $FeCl_3$ carbon that formed during the high temperature MP process. This conversion process is poorly reversible and no clear dQ/dV feature is observed for the reverse process on charge. The $MnCl_2$ anode exhibits a similar voltage plateau (**Figure S4c**) and dQ/dV feature (**Figure S4d**) in the range of 0.8-0.5 V vs. Li/Li$^+$ on discharge, and we thus propose a similar mechanism of $MnCl_2$ conversion into LiCl, whereby $MnCl_2$ + 2Li$^+$ + 2e$^-$ → Mn + 2LiCl. Interestingly, this conversion reaction appears to be more reversible than in the case of the $FeCl_3$ carbon, with a corresponding dQ/dV feature at ~1.2 V vs. Li/Li$^+$ observed on charge. Conversion reactions involving electrochemically-active salts are problematic as they likely contribute to the large irreversibilities observed during the first discharge-charge cycle for the pristine $FeCl_3$ and $MnCl_2$ carbons, and they also consume a significant amount of Li. We find that the reversible capacity and average working potential of the $FeCl_3$ carbon consistently increases with each additional washing step. Because the working potential of this carbon is generally higher than the potentials of graphite (solid line) and hard carbon (dashed line) reference electrodes, the possibility of having more accessible graphitic or disordered carbon domains upon washing can be eliminated. We instead attribute the observed increase in the working potential and reversible capacity upon washing to partial oxidation of residual $FeCl_3$ catalytic salt impurities, leading to a more reversible conversion process with Li$^+$ ions. Indeed, a comparison of the dQ/dV plots obtained on charge for the pristine, water washed and water + $NO_2BF_4$ washed $FeCl_3$ carbons (see **Figure S4e**) shows an increase in the intensity of high voltage dQ/dV features with washing; features that likely correspond to the reversed conversion process on charge. In contrast, the reversible capacity of the MnCl$_2$ carbon gradually decreases with further washing, and the decrease in its Li extraction potential after a water + NO$_2$BF$_4$ wash could indicate reduced side reactions between Li$^+$ ions and the MnCl$_2$ salt. As shown in **Figure S4f**, the dQ/dV peak at ~1.2 V on charge shifts to higher potentials but does not change in intensity after a hot water wash, while an additional NO$_2$BF$_4$ wash both shifts it to even higher potential and reduces its intensity, consistent with gradual oxidation and removal of residual catalytic salt.

Overall, our results suggest that a hot water wash of pyrolyzed carbons is beneficial to reduce irreversible electrochemical reactions consuming Li, especially those between redox-active FeCl$_3$/MnCl$_2$ salts and Li$^+$ ions. The additional NO$_2$BF$_4$ washing step only seems to remove a small amount of the MnCl$_2$ salt from the carbon and does not seem to improve the performance of the other carbons. Hence, it is not cost-effective for large-scale applications.

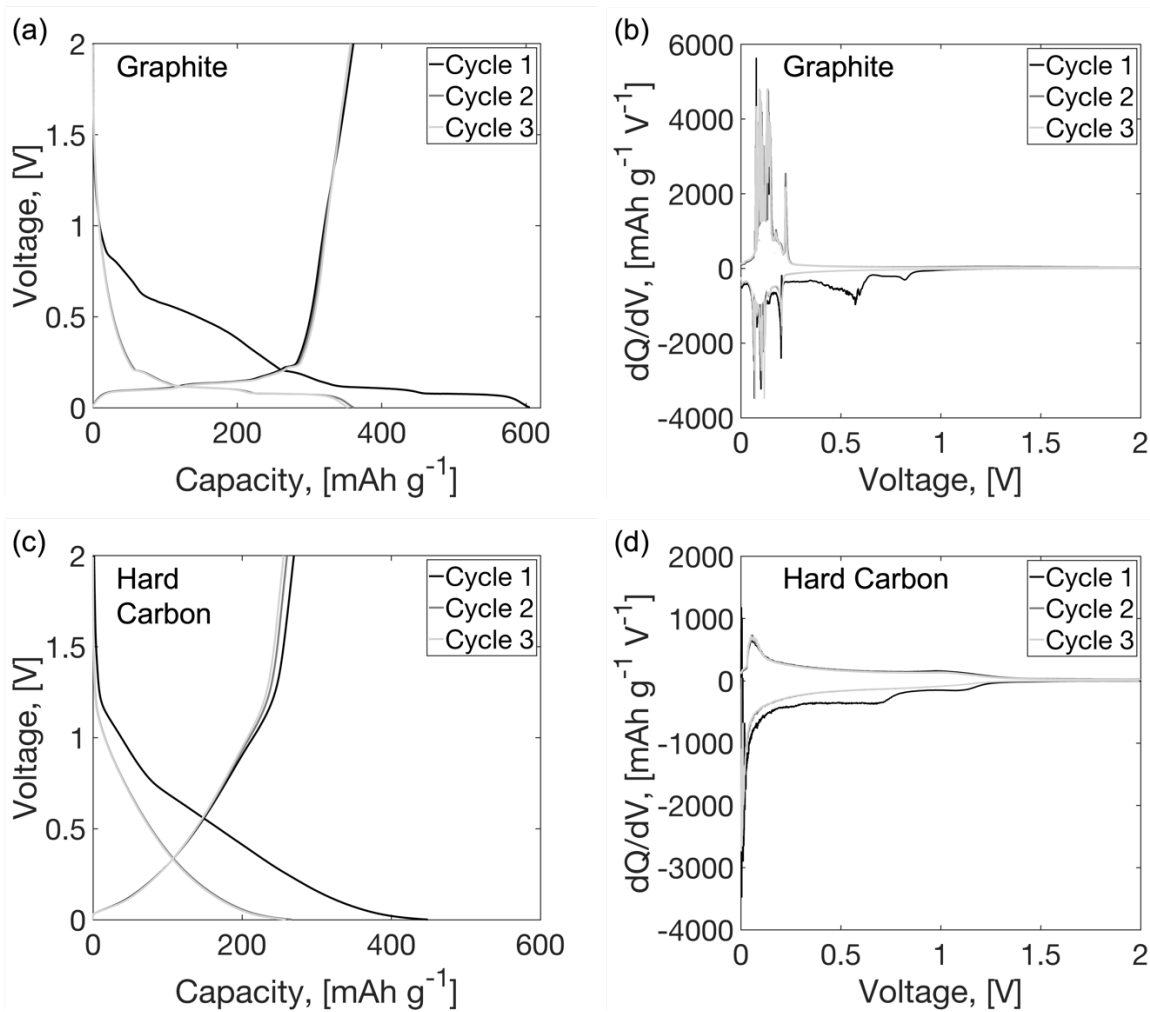

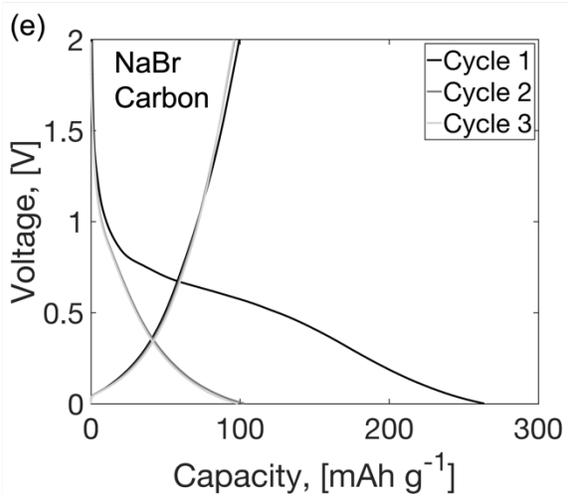
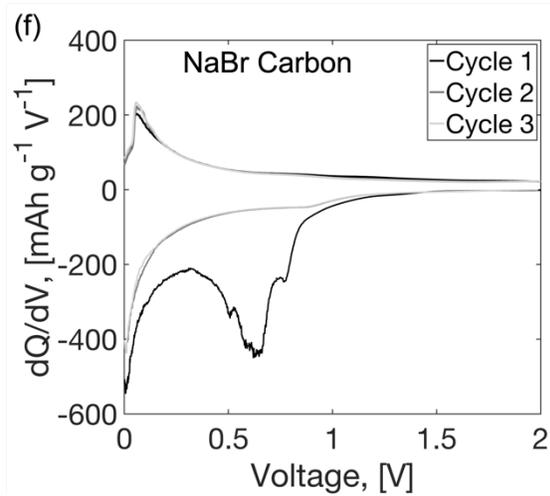
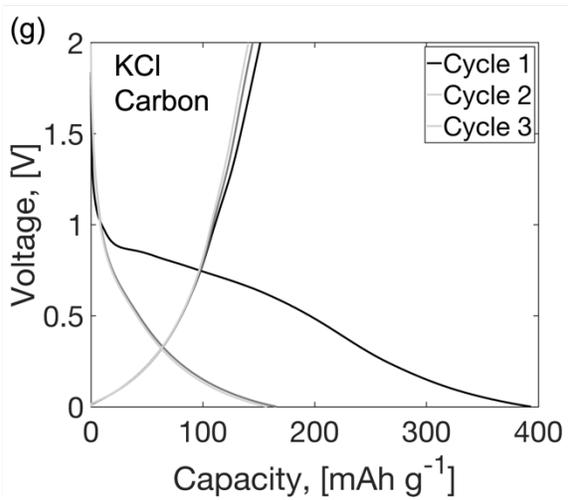
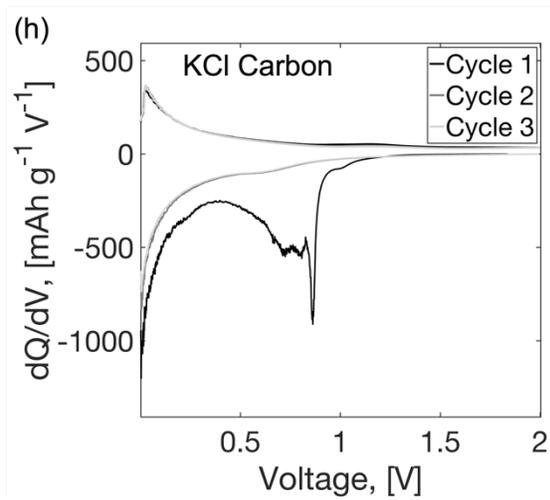
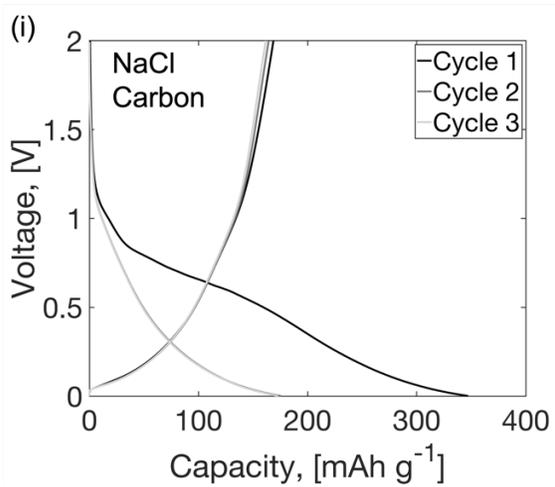
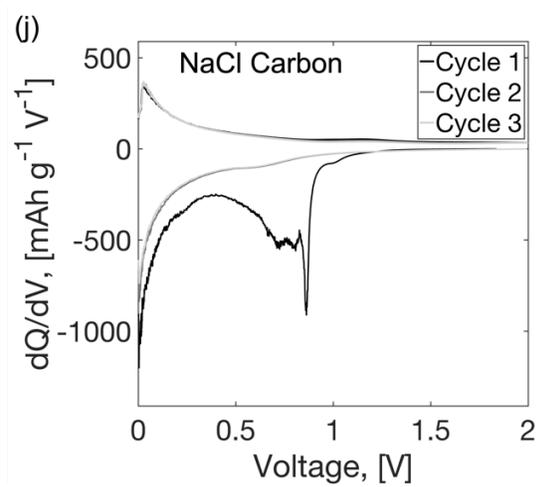

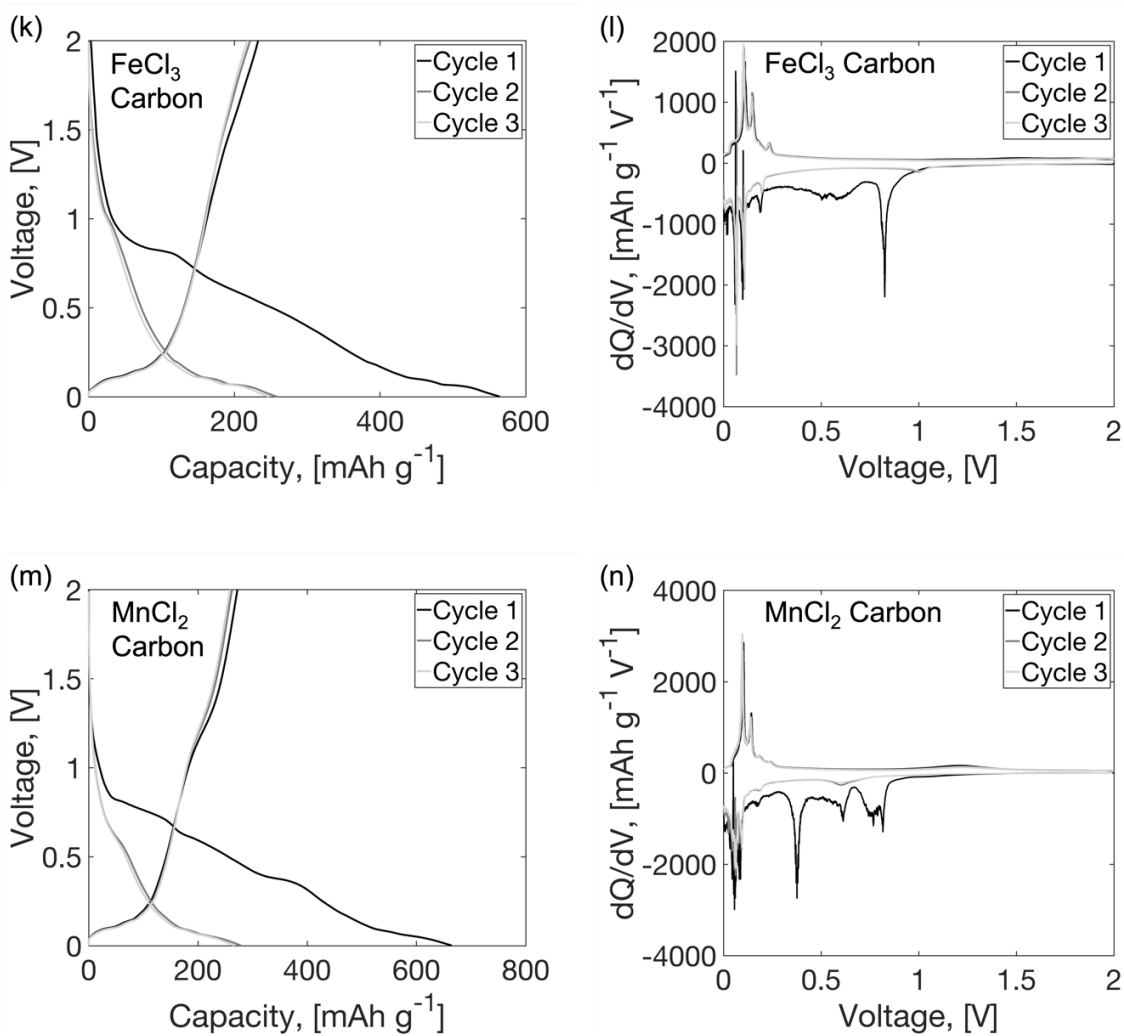

**Figure S5.** Electrochemical properties over the first 3 discharge-charge cycles for various carbon electrodes dried after a hot water wash tested in Li-ion half cells over a 0.001 to 2 V vs. Li/Li$^+$ potential range and at a rate of C/20 (based on graphite). The voltage vs. capacity curves are shown in: a) Graphite, c) hard carbon, e) NaBr carbon, g) KCl carbon, i) NaCl carbon, k) FeCl$_3$ carbon,, and m) MnCl$_2$ carbon. The differential capacity (dQ/dV) plots are shown in: b) Graphite, d) hard carbon, f) NaBr carbon, h) KCl carbon, j) NaCl carbon, l) FeCl$_3$ carbon, and n) MnCl$_2$ carbon.

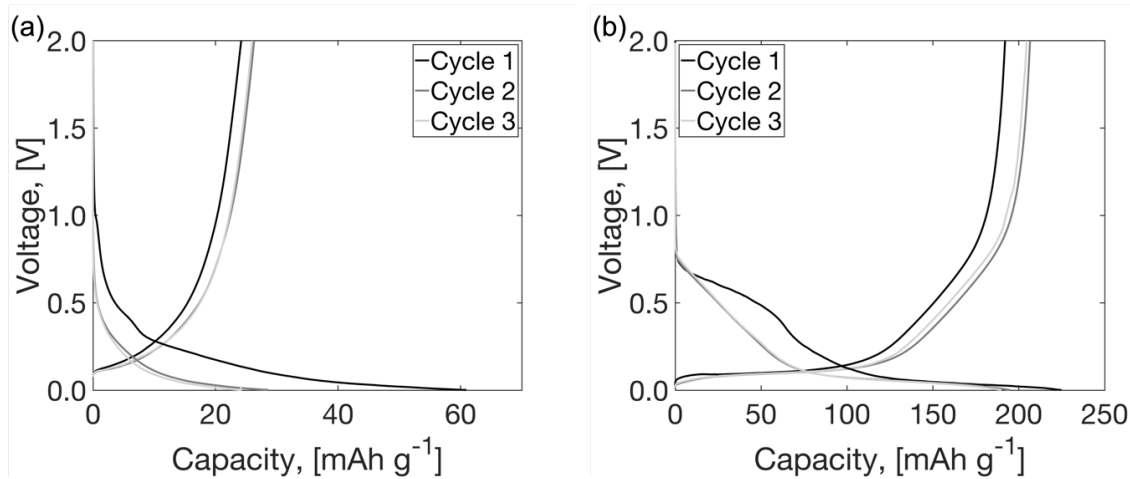

**Figure S6.** Voltage profiles in Na-ion half cells of a) water washed KCl carbon and b) commercial hard carbon electrodes cycled at a rate of C/20 (based on graphite).

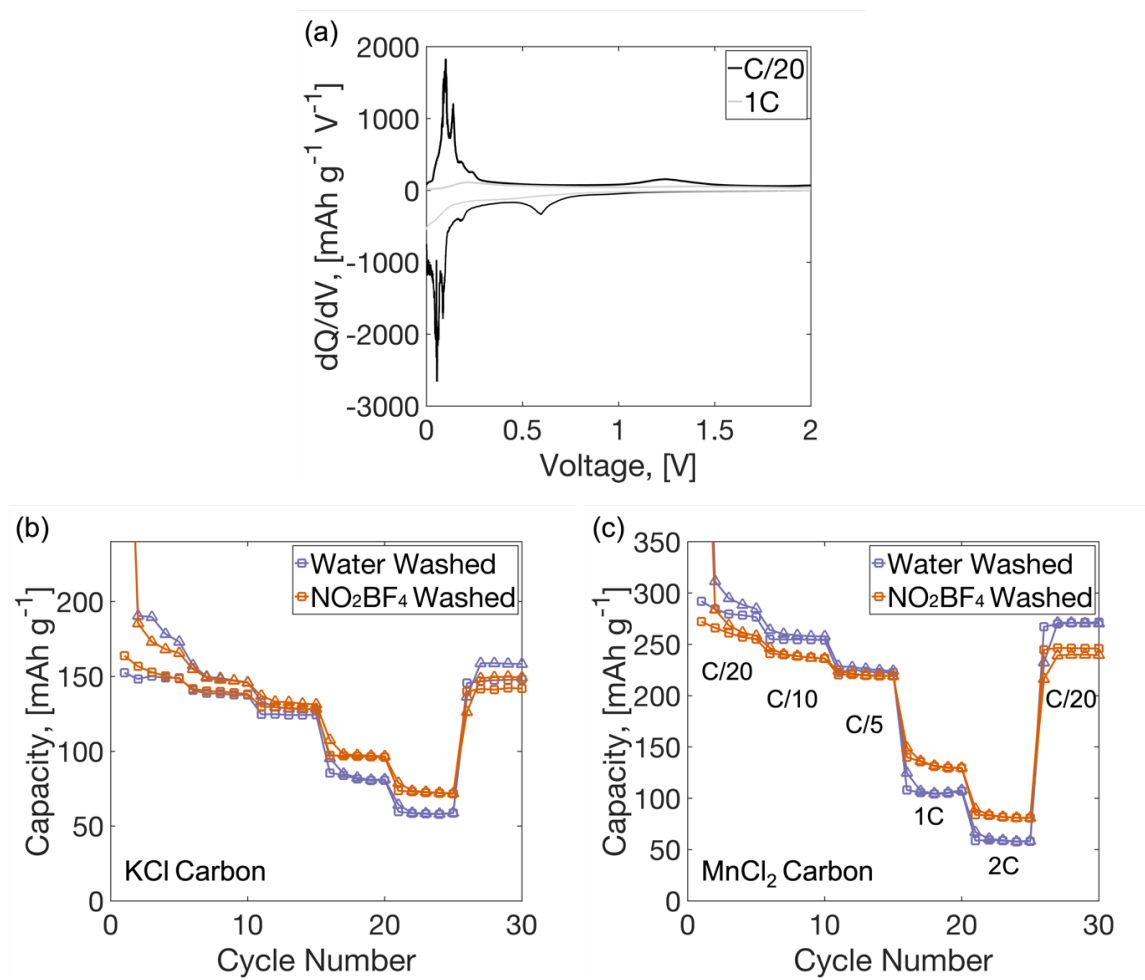

**Figure S7.** a) dQ/dV plot on discharge of the MnCl$_2$ carbon electrode cycled at C/20 and 1C. Capacity observed for the b) KCl carbon and c) MnCl$_2$ carbon electrodes after a water (purple) and a water + NO$_2$BF$_4$ (orange) wash upon step increases in cycling rate from C/20 to 2C and back to C/20 over 30 cycles.